\documentclass{easychair}
\usepackage{makeidx}

\newcommand{\xf}[1]{Figure~\ref{#1}}

\newcommand{\xs}[1]{Section~\ref{#1}}

\newcommand{\tool}[1]{\texttt{#1}\index{Tools!#1}}

\newcommand{\api}[1]{\texttt{#1}\index{API!#1}}

\newcommand{\lucidL}[1]{{$\mathit{Lucid}$}($L$) }

		{}

\def\myvert{\raise 2.27pt \hbox{\vrule depth 0pt height 8pt width 0.2mm}}
\def\myarrow{\hspace*{0.43mm}%
             \raise 2.29pt\hbox{\vrule depth 0pt height 8pt width 0.16mm}%
             \hspace*{-0.32mm}%
             $\longrightarrow$
             \ %
             }

\newcommand{\BigSwitchOne}{BigSwitch1}
\newcommand{\BigSwitchTwo}{BigSwitch2}

\newcommand{\catalyst}[1]{Catalyst #1\index{Switches!Catalyst #1}}
\newcommand{\swTNFZ}{170 \catalyst{2950}}
\newcommand{\swTNXZ}{10 \catalyst{2960}}
\newcommand{\swTNSZ}{10 \catalyst{2970}}
\newcommand{\swTNTF}{10 \catalyst{2924}}
\newcommand{\swTFFZ}{10 \catalyst{3550}}
\newcommand{\swTFFE}{10 \catalyst{3548}}

\newcommand{\wclosets}{65}
\newcommand{\netclients}{4800}
\newcommand{\manageddesktops}{1100}
\newcommand{\jackpatches}{5800}

\newcommand{\vlans}{200}
\newcommand{\fwrulesIN}{730}
\newcommand{\fwrulesEX}{450}

\newcommand{\dbentities}{80}
\newcommand{\scripts}{60}

\newcommand{\cisco}{Cisco}

\makeindex

\begin{document}

\title{Multifaceted Faculty Network Design and Management: Practice and Experience Report
\footnote{A \href{http://dx.doi.org/10.1145/1992896.1992916}{short version} was presented at C3S2E 2011.}}
\titlerunning{Multifaceted Faculty Network Design and Management}

\authorrunning{Michael J. Assels et al.} 

\author{
Michael J. Assels$^1$, Dana Echtner$^2$,
Michael Spanner$^1$, Serguei A. Mokhov$^1$,\\Fran\c{c}ois Carri\`{e}re$^1$, Manny Taveroff$^1$\\
$^1$Faculty of Engineering and Computer Science\\
Concordia University, Montreal, Quebec, H3G 1M8, Canada\\
\url{{mjassels,spanner,serguei,frank,manny}@encs.concordia.ca}\\
\and
$^2$Faculty of Computer Science\\
Dalhousie University, Halifax, Nova Scotia, B3H 1W5, Canada\\
\url{dana.echtner@dal.ca}\\
(Formerly at Concordia)
}

\maketitle              

\begin{abstract}
We report on our experience on multidimensional aspects of our
faculty's network design and management, including some unique aspects such
as campus-wide VLANs and ghosting, security and monitoring, switching and routing,
and others. We outline a historical perspective on certain research,
design, and development decisions and discuss the network topology, its scalability,
and management in detail; the services our network provides, and its evolution.
We overview the security aspects of the management as well
as data management and automation and the use of the data by other
members of the IT group in the faculty.
\end{abstract}

\tableofcontents
\listoffigures

\section{Introduction}
\label{sect:intro}

We present a number of the aspects we designed in the network infrastructure
at the
Faculty of Engineering and Computer Science (ENCS), Concordia University,
Montreal, Quebec, Canada.
The experience report covers about 10 years of
design, development, deployment, and change management 2001--2010. 
We first very briefly touch on some historical notes necessitating the
presented design and its evolution. Then we highlight the key problems and motivation in \xs{sect:problems}
and proposed solutions to those problems in \xs{sect:solutions}.

\subsection{Brief historical notes}
\label{sect:brief-history}

We are reporting on the evolution of a functioning network as it
grew through {\em ad hoc} responses to problems over more than ten
years, not on a careful design of a freshly implemented network.
When the IT personnel under various departments such as CSE, ECE, MIE, BCEE, etc.
were merged under the umbrella of the Faculty of ENCS, the smaller departmental
networks and their management infrastructure had to be streamlined.
These mergers saw
the network grow approximately 4-fold (this represents about $1/3$ of the entire University's
network). New policies, management strategies, security\index{security}
considerations, reliability, and scalability\index{scalability} had to be addressed, especially, since, at the
time of the merger the networks were spanning more than 10 geographically
distributed buildings. The present network is comprised of approximately {\netclients}
network clients
that include desktops, servers, and other networked devices.

\subsection{Problems and challenges}
\label{sect:problems}

As we report on an evolutionary process, with motivation for change sometimes being
political (as in the case of integrating multiple departments),
sometimes technical (as in the case of needing to run multiple
simultaneous ghosting\index{ghosting} sessions without disturbing the global
network), and sometimes managerial (as in the case of our
need for a network database\index{database} to track locations of equipment
in near-real time).
As a result, several key issues emerged:

\begin{itemize}
	\item 
Security\index{security}.

	\item 
Scalability\index{scalability} and management.

	\item 
Accretion of multiple departments with different setups.

	\item 
Trusted subnets (analyst-managed) and untrusted subnets (user-managed).

	\item 
Ghosting\index{ghosting} (OS image cloning).

	\item 
Cooperation with other IT subgroups.

	\item 
Connectivity and reliability.
\end{itemize}

\subsection{Proposed solutions}
\label{sect:solutions}

The solutions deployed to address the challenges are summarised here:

\begin{itemize}
	\item 
Campus-wide VLANs\index{VLAN!Campus-wide}. The novelty in our design in here included, which
nobody else to our knowledge had ghosting\index{ghosting} VLANs\index{VLAN!ghosting} capable of ``imitating''
the global network in multicast\index{multicast} sessions without adversely
affecting normal operations.

	\item 
Extensive scripting\index{scripting} and database\index{database} support
for management and monitoring\index{monitoring}.

	\item 
Spanning tree\index{spanning tree} setup.

	\item 
Development of a set of audited tools and shells to allow
sister group to access the data and perform
simple networking tasks.

	\item 
Extensive internal and external firewall\index{firewall} design.
\end{itemize}

\subsection{Organisation}

What follows are some details on the solutions, best practices, new
challenges and future work.
In \xs{sect:campus-vlans} the ideas and their realization are described
behind the notion of campus-wide VLANs\index{VLAN!Campus-wide}. 
In \xs{sect:security} we describe security\index{security} aspects of the network design and
our conservative approach
which allows for more flexibility.
In \xs{sect:switching} are some switchfarm\index{switchfarm} deployment details,
whose configuration allows for spanning tree\index{spanning tree} to function.
In \xs{sect:routing} are the design considerations of our routing\index{routing} setup
taking into consideration the security\index{security} and campus-wide-VLAN aspects.
In \xs{sect:services} we describe the type of services\index{services} we provide with our network
to our faculty's community.
We provide some concluding remarks and statistical summary in \xs{sect:conclusion}.

\section{Switching}
\label{sect:switching}

We describe some our switching topology\index{switchfarm} and configuration
details
to highlight key points of interest.

\subsection{Configuration}
\label{sect:switch-configuration}

We maintain a common central switch configuration profile, which is
used when deploying a new switch. In that profile, we typically configure the
DNS settings, switch logging\index{logging}, and NTP hosts (for automation and monitoring\index{monitoring}
of switch logs).
Standard procedures are in place for switch
additions, removals, moves, etc. as far as the configuration is concerned, and
allowing the spanning tree protocol\index{spanning tree}\index{Protocols!spanning tree} to function properly.

\subsection{Spanning tree protocol}
\label{sect:spanning-tree}
\index{spanning tree}\index{Protocols!spanning tree}

Following the mathematical notion of spanning tree (connected undirected graph), the
corresponding switching protocol \cite{perlman-spanning-tree-1985,wiki:spanning-tree-protocol,cisco-kb-stp}
(STP)\index{spanning tree}\index{Protocols!spanning tree} is fully implemented
in our network that covers pretty much the entire
switch farm.
We have a single spanning tree covering all our switches,
that facilitates failover as well as redundancy.
The main point of the protocol\index{Protocols!spanning tree} is to have connectivity coverage
to all the leaf nodes without creating an accidental loop directly
or indirectly ``shortcircuiting'' any two or more switch ports
followed by the shutdown of the tree. To keep it simpler, we don't have a mesh of switches,
but we make the extensive use of port channels to maintain
redundancy and most of our VLANs are trunked on all switches.
Tree structure of a connected element is a triangle (with one
``corner'' ``broken'', so it is still an acyclic graph).
One of the ``access'' switches in the triangle is connected to another to form a pair
via a gigabit port (via GBICs), and other ports of the corresponding switches
go to the core stack described below with one of the ports configured
to be in blocking mode to prevent looping. Breaking any of the links would
not result in a loss of connectivity as the redundant path would be followed.

Two major buildings have a stack of {\cisco} Catalyst 3750 switches.
The main one, is the 9-element stack of {\BigSwitchOne}, is now arranged in a ring
(see \xf{fig:stack-ring}).
\begin{figure}[htpb!]
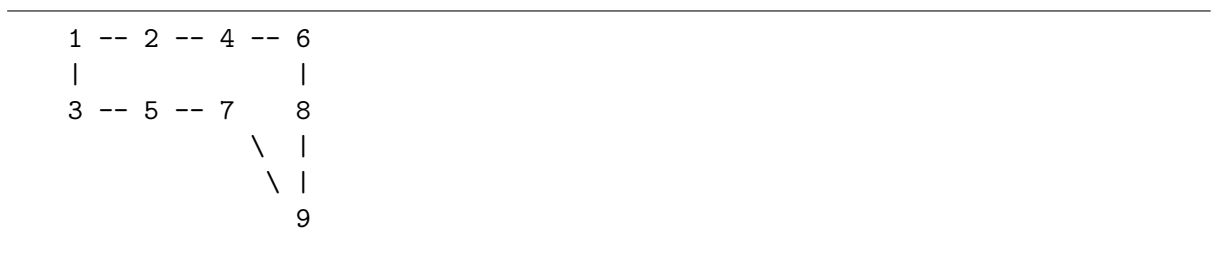
%
\hrule
\begin{verbatim}

    1 -- 2 -- 4 -- 6
    |              |
    3 -- 5 -- 7    8
                \  |
                 \ |
                   9
\end{verbatim}
\hrule
\caption{Ring connectivity in the main core stack of 3750s}%
\label{fig:stack-ring}%
\end{figure}
Elements in the upper row are powered by one UPS, and
elements in the lower row (and 9) are powered by another.
We have taken care to connect all the spanning-tree-paired switches
to corresponding port numbers on vertically matched stack elements;
e.g., a switch $A11$ (1/0/15) (stack element 1, port 15) is connected to another switch $A12$ (3/0/15), and
a switch $A13$ (2/0/9) is connected to another $A14$ (5/0/9).  That way,
even if we lose a UPS in Room A, we still have full switch
connectivity.
To elaborate, an access switch $A11$ in wiring closet $Wc1$ is connected through
one of its gigabit ports, say, to stack element 1, port 15; its
other gigabit port is connected to its ``closet-mate'' $A12$, which
is itself connected to stack element 3, port 15, thus forming
a three point loop for the spanning tree protocol\index{spanning tree}\index{Protocols!spanning tree} to handle.
In normal operation, spanning tree will very quickly decide
to block the direct connection between $A11$ and $A12$, sending
all traffic through the core stack. If either uplink to the
core stack is lost, or if one of the stack elements 1 or 3
is lost, spanning tree will quickly unblock the $A11-A12$ link,
reestablishing full connectivity. If one of the UPSes in the
network operations center should fail, \xf{fig:stack-ring} shows that
only one of the stack elements 1 and 3 will be lost, and once
again, spanning tree will quickly unblock the $A11-A12$ link,
reestablishing full connectivity.

When connecting a non-switch device such as a server, we
leave its ``paired port'' empty, but we preconfigure
it so that the device's cable can just be moved to the
paired port.
For the partnerless switches we give
double uplinks to both paired ports.
Switch \#9 is a spare. If switch \#n fails, \#9 can 
be reconfigured to take over from it.
The schematic of the described connectivity is illustrated
in \xf{fig:before} (current configuration) where {\BigSwitchOne}
and {\BigSwitchTwo} comprise the ``Central stack A'' and ``B''.

\begin{figure}[htp!]
	\centering
	\includegraphics[width=\columnwidth]{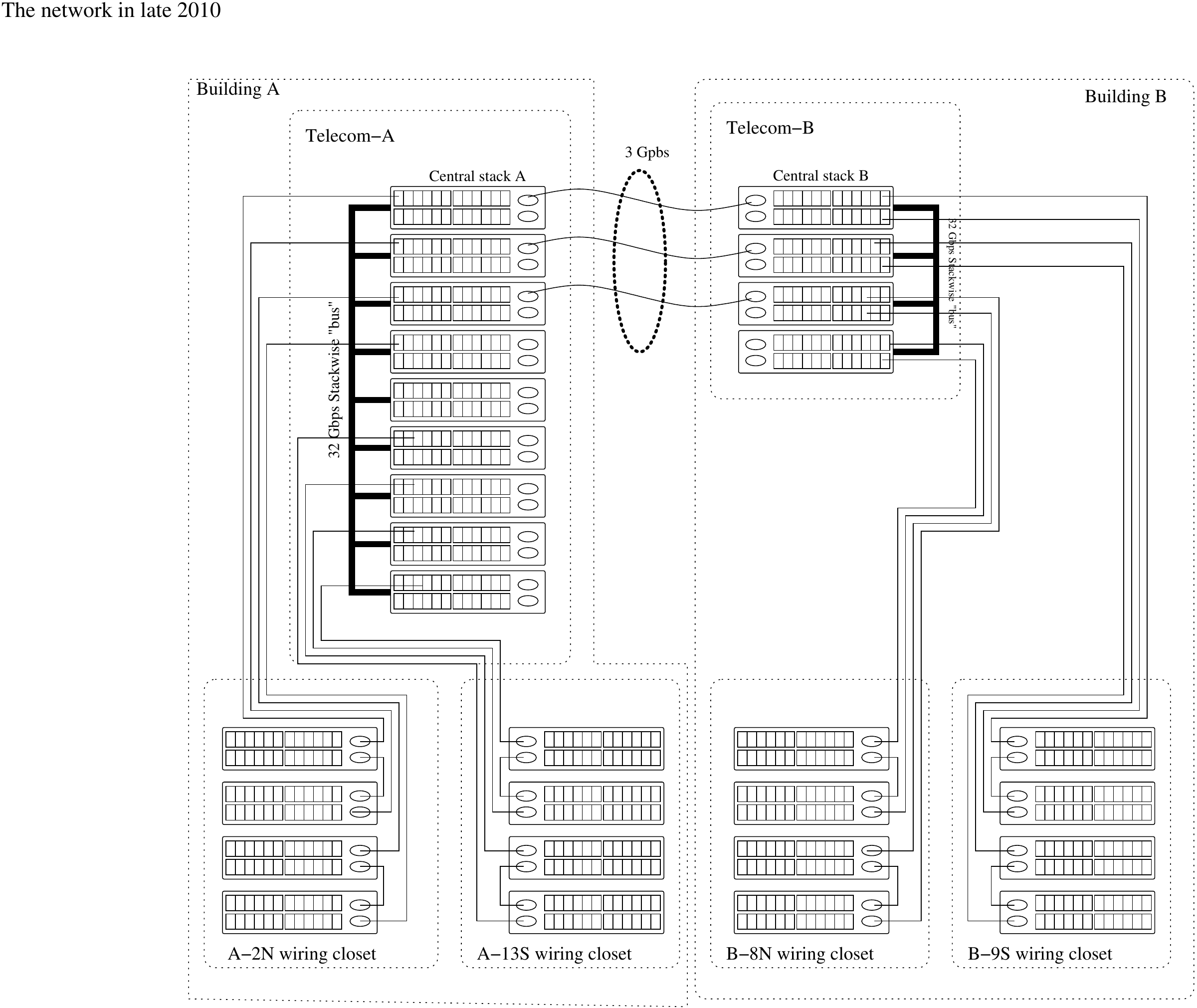}
	\caption{Network leading up to 2010}
	\label{fig:before}
\end{figure}

\begin{figure}[htp!]
	\centering
	\includegraphics[width=\columnwidth]{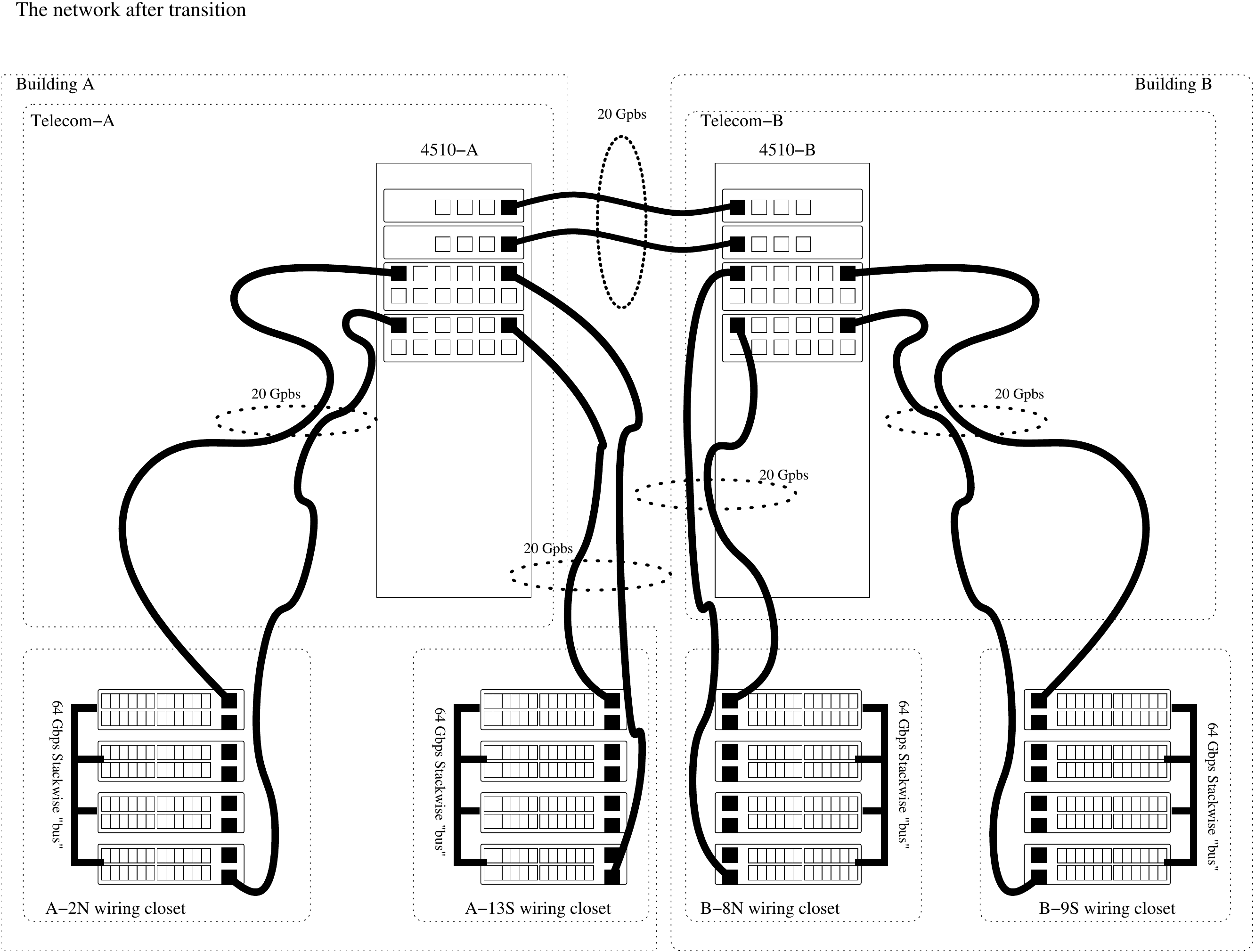}
	\caption{Planned major core network changes to upgrade}
	\label{fig:after}
\end{figure}

\begin{figure}[htp!]
	\centering
	\includegraphics[width=\columnwidth]{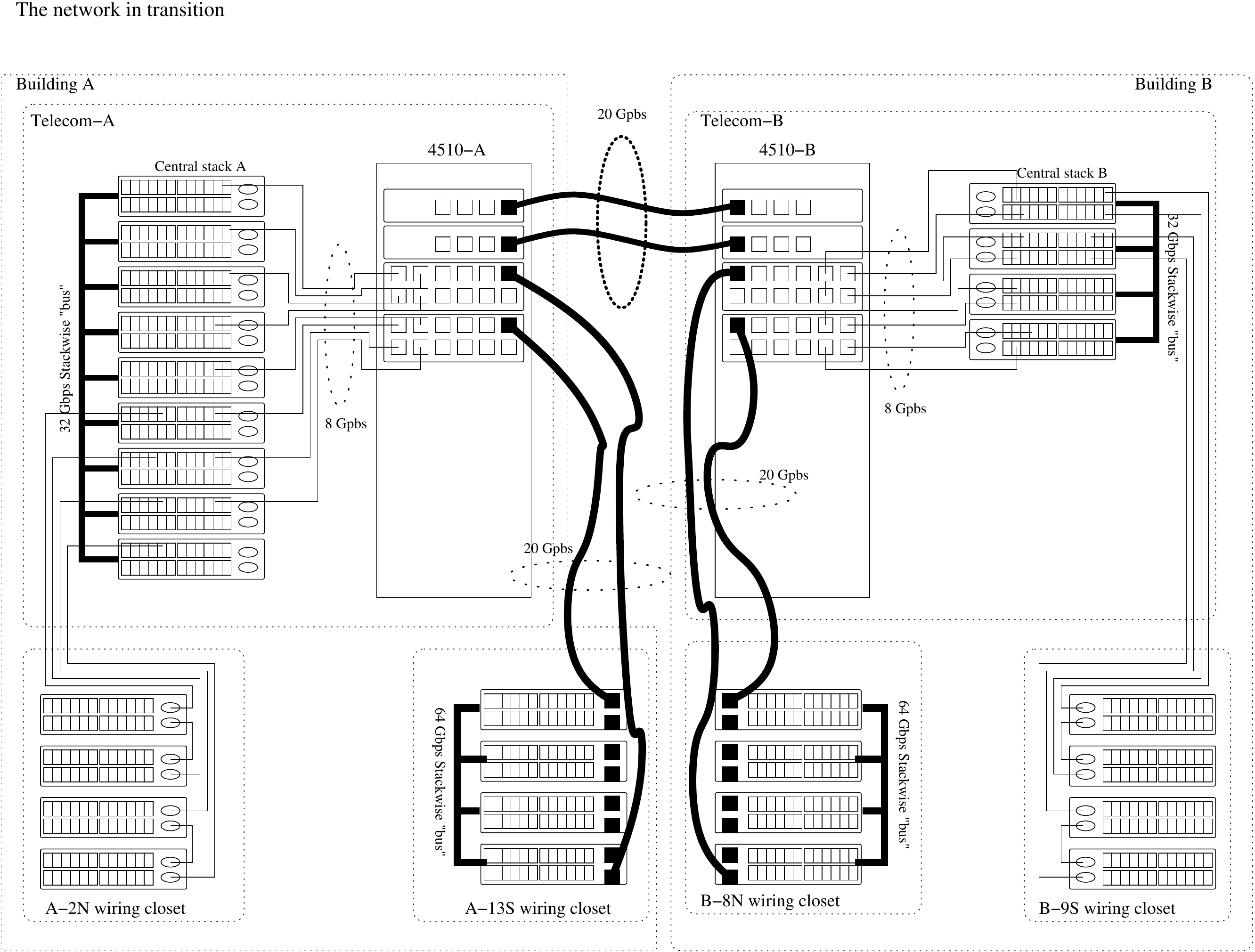}
	\caption{Transitional/migration setup}
	\label{fig:during}
\end{figure}

\subsection{Campus-wide VLANs}
\label{sect:campus-vlans}

We diverge from the classical {\cisco} network setup~\cite{cisco-lan-switching-99}, where
the backbone of the network consists of the core switches connected to distribution
and access switches per floor with access switches for individual end nodes (computers, etc.).
This would inhibit our
current design's ability to setup any computer on
any VLAN anywhere on campus instantly. This is where our innovation comes in.
When we look at the standard {\cisco} network setup, we could see
that our configuration draws some similarities.
These similarities can be seen if we look at
the upstream provider for our network's
connectivity
as the core network.
Then, our core {\BigSwitchOne} and {\BigSwitchTwo}
switch stacks 
(that correspond to the knots in \xf{fig:netmap})
as well as
the smaller gigabit switches act as distribution switches, and
the spanning-tree-paired switches would play the role of the access switches.

Initially, we have not implemented Layer 3 connectivity on our
switchfarm\index{switchfarm} opting instead to have the routing\index{routing} and
firewalling\index{firewall} along with NATing\index{NAT} performed by a redundant set
of Linux boxes (one in each building via
the core fibre-interconnected switches, see \xf{fig:before}).
This design was chosen because of constraints imposed
by the ghosting\index{ghosting} requirement (``ghosting'' as in cloning desktop
operating systems' images using the Symantec's Ghost software
that allows multicast distribution of the disk images to multiple
clients simultaneously) as well as the need for
geographically dispersed VLANs for faculty members.

\subsubsection{Multiple ghosting sessions.}
\label{sect:mult-ghosting-sessions}

The Desktop group for some duration of time was performing 10 simultaneous
ghosting\index{ghosting} sessions across $\approx {\manageddesktops}$ analyst-managed computers primarily
spread across two buildings (earlier there were more buildings to cover before
the Faculty got consolidated). A single person was able to ghost a floor of
$\approx 300$ computers at a time. The entire ghosting\index{ghosting} season would last 3-5 days
to cover the majority of the lab computers. The ghost image sizes varied
for 10GB for a Linux partition and about 20GB for a Windows partition.
To accomplish such a feat without significant interruption of the network
service to the rest of the regular users during the ghosting\index{ghosting} season
the Desktop group required a network service that could support all this.
One way to do this was allocating a VLAN per Desktop group member, where
each Desktop member also has their own ghosting\index{ghosting} server -- two ghosting
servers did not play well together in the same VLAN -- as well
as a ghost router, which is a copy of one of our main router pairs
along with the DHCP server copy to work in the ghost VLANs. In order to
ghost any arbitrary number of computers normally residing in their respective
VLANs and potentially across buildings, a Desktop group member would move
the switch ports of all those machines (via the scripting and shell services\index{services}
we implemented in-house, see \xs{sect:scripting}) to their own ghosting VLAN\index{VLAN!ghosting}
in batch, and can do the entire ghosting\index{ghosting} session of this arbitrary set of
machines without disruption to the main networking services\index{services} anywhere else
in the Faculty.

This could only be possible with campus-wide VLANs trunked on all switches
and not a traditional {\cisco}-world setup \cite{cisco-lan-switching-99,cisco-catalyst-2950-2003}.

\subsubsection{Custom VLANs for faculty members.}
\label{sect:vlans-faculty}

To secure and seclude research groups or individual Faculty members for the
computers they purchase for them, their students and affiliates to do
the research, share files, etc. they are usually allocated a VLAN and a subnet
of a required size depending on their needs.
This isolates them from potential attacks from other groups and subnets
while permitting
network printing
and file sharing they need. It also happens
in some cases some professors or research groups span their presence across
buildings as well necessitating the VLANs being available anywhere on the
campus at any point in time.

\section{Routing}
\label{sect:routing}
\index{routing}

This report covers essentially the modus operandi between circa 2001 to 2010,
where all our switchfarm\index{switchfarm} was Layer-2\index{switchfarm!Layer 2} only. The routing\index{routing}, as a result, has been
performed on pairs of redundant Linux boxes geographically spread across
buildings, and is comprised of an internal pair (among our own subnetworks) and
an external pair (the edge to the outside). The routers
impose the firewall\index{firewall} rules designed for different purposes (see \xs{sect:firewalls}).
The routing\index{routing} supports all the required virtual interfaces for the VLANs
present on the network and the maintenance of the routing\index{routing} tables.

\paragraph{Connection tracking.}
\label{sect:conntrack}

The routers employ connection tracking for failover between buildings
without loss of sessions where the corresponding tables are saved and
reloaded from one end (primary) to another (secondary) in case of a failure of the primary or reboot
of the primary, automatically.

\section{Services}
\label{sect:services}
\index{services}

Our network design, besides providing the usual ``access to the Internet''
is geared towards services\index{services} for the students, Faculty and staff.
There are all kinds of specialised needs for printing, filer sharing,
servers, research equipment and groups including research
graphics cluster, as well as other research clusters in parallel
computing, software security\index{security}, service-oriented architectures,
language engineering, networking research, database\index{database} engineering,
genomics, audio-visual, pattern recognition, and others.

The services\index{services} to our own IT subgroups are provided in the form
of the network database\index{database} data IT staff require for their daily activities.
The data are provided via access tools (see \xs{sect:scripting}), which enable IT staff to perform
a restricted set of networking tasks in order to improve our scalability\index{scalability}
as an IT group.

\paragraph{AL.}
\label{sect:al}

A now historical service to support {\em authenticated laptop} (AL) connections
by the users with a valid account in the Faculty and a dynamic IP within
a range set aside for AL.
AL was an important service to our clientele.  While it
has significant security\index{security} considerations, as nearly any our service does, but
AL is a service more than a security\index{security} problem.
Most of the switch ports were configured with \api{dot1x}, so potentially
any unused patched jack would be in the AL state so a potential user
could use it to authenticate themselves to be allowed to connect further.
At the present, AL replaced by the allocation of static IP addresses on
our network to the users that make a proper request and are not using
wireless connections. The replacement was necessitated by the decision of
the parent IT unit to abandon the use of proprietary software for Windows
platforms since that platform does not have a native implementation
of two-factor 802.1X\index{802.1X}\index{Protocols!802.1X} authentication built-in at the time of this writing,
and the proprietary software licensing terms became unbearable. While
the AL service could still support Linux and MacOS X clients, their number,
while increasing, is still small comparative to the Windows clients.

\section{Security}
\label{sect:security}
\index{security}

In the AITS group, IT security spans aspects covering several subgroups besides networking, including
system administration, desktop operations, faculty information systems, and user
services\index{services}. Our responsibly is to ensure a sane and secure networking environment
to the entire Faculty and guard the integrity and availability of our backbone
core and the leaf nodes
\cite{tao-net-sec-2005,conu-sec-anne-mja-1995,pervasive-trust-ngns,rubin-white-hat-arsenal-2001}.

This covers the network bandwidth monitoring\index{monitoring!bandwidth}, health of the switch farm, routing\index{routing}
equipment, and rogue DHCP servers monitoring\index{monitoring!rogue DHCP} as well as desktop monitoring for
infections\index{monitoring!infections} and vulnerabilities\index{monitoring!vulnerabilities} (with quarantine), potential violations by
MAC spoofing attacks, port scanning, illegal peer-to-peer activities,
and equipment moves.

\subsection{Firewalls}
\label{sect:firewalls}
\index{firewall}

We employ two pairs of redundant firewalls across two buildings
implemented using \tool{iptables} in Scientific Linux environment
\cite{linux-firewalls-ids-2007,linux-iptables-pocket-2004,linux-firewalls-2006}
that provide a number of internal and external services\index{services}, routing\index{routing}, NATing\index{NAT}, logging\index{logging}
and monitoring\index{monitoring}. Both firewalls\index{firewall} total in more than 1000 rules. Both firewalls
employ conservative deny-all policy and allow only required services\index{services}.

{\em Why Scientific Linux and iptables?}
While any Linux distribution would be fine, we traditionally used RedHat derivatives
such as RHEL, Fedora, and now Scientific Linux as it is a widely deployed and entreprise level
operating system. The primary justification for \tool{iptables}
(aside from the cost and being one of the most widely deployed industry standard firewalls\index{firewall}) is that
\tool{iptables} allows unrestricted branching of the rule ``tree'', something that's important to the
task of securely managing traffic amongst more than 150 subnets
offering different services\index{services} and presenting widely varying
security\index{security} problems. {\cisco} ACLs e.g. are simply not up to the task.
Additionally, when we started, {\cisco} ASA didn't exist, but \tool{ipchains} (precursor of \tool{iptables})
did. Moreover, cost was an issue.  A ``layer 2'' {\cisco} network
with ``layer 3'' handled by commodity hardware and free and open-source software
was at the same time much cheaper and more flexible than any available
alternative.

The internal firewall\index{firewall!internal} provides or denies access to or from our IT's
core servers managed by the system administration group and desktop groups
to allow printing, connectivity to the NetApp filers, web applications and
the like. Misbehaving hosts are also placed in quarantine on the internal
firewall\index{firewall!internal}
(e.g. in \xf{fig:typical-quarantine-entry}
what we configure to
place on quarantine and the resulting \tool{iptables} rules are
in \xf{fig:typical-quarantine-firewall-rules} compiled with a \tool{make}
\cite{gmake}
marking at pre-routing\index{routing} with a special token; and anything marked with that token
at the forward chain is only allowed accessing permitted patch and antivirus sites).

\begin{figure}[htpb!]
\hrule
\scriptsize
\begin{verbatim}

...
# admnusr 2010-11-03 XXX.XXX.XXX.134 8 OS[Windows 5.1] MS06-040 VULNERABLE
sickhost.domain
...
\end{verbatim}
\normalsize
\hrule
\caption{Typical quarantine entry}
\label{fig:typical-quarantine-entry}
\end{figure}

\begin{figure*}[htpb!]
\hrule
\scriptsize
\begin{verbatim}

Chain PREROUTING (policy ACCEPT)
target      prot opt source          destination
MARK        all  --  sickhost.domain anywhere        MARK set 0x2
...
Chain FORWARD (policy ACCEPT)
target      prot opt source          destination
qrntine     all  --  anywhere        anywhere        MARK match 0x2
...
Chain qrntine (1 references)
target      prot opt source          destination
ACCEPT      all  --  anywhere        XXX.XXX.XXX.0/24
ACCEPT      all  --  anywhere        anywhere
ACCEPT      tcp  --  anywhere        antivirus-site1.domain tcp dpt:1234
ACCEPT      tcp  --  anywhere        antivirus-site1.domain tcp dpt:1234
patchSites  all  --  anywhere        anywhere
REJECT      tcp  --  anywhere        anywhere        reject-with tcp-reset
REJECT      all  --  anywhere        anywhere        reject-with icmp-port-unreachable
\end{verbatim}
\normalsize
\hrule
\caption{Typical quarantine set of firewall rules}
\label{fig:typical-quarantine-firewall-rules}
\end{figure*}

The external firewall\index{firewall!external} deals with the connectivity to and from outside of
our administrative domain allowing certain incoming connections such as web,
\tool{ssh}, remote desktop, licensing, and printing services\index{services} from the University's wireless,
and the like to an authorised list of hosts. Offending phishing and
other external sites are blocked at this firewall\index{firewall!external}. NATing\index{NAT} from the private
space as well as bandwidth restriction enforcement take place here as well.

The list of blocked internal machines and external sites are provided
via a web page to the rest of IT staff, primarily the service desk
to help to deal with phishing attacks and the related inquiries.

The rules are processed from a source script by a bash script in order
to prepare an \tool{iptables-restore} loadable file that pre-resolves
the DNS entries to speed up processing and deter errors as well as loads
the rulesets automically to avoid race conditions, etc. -- in an approach
that was similarly presented in \cite{iptables-atomic-changes-2010}.

\subsection{Switch port violations}
\label{sect:swpvios}

We also monitor switch port violations ({\em swpvios}).
Most of the ports are configured to be bound to a fixed number of
known MAC addresses (usually one). When users attempt to move computers
around or plug-in an unknown piece of networked equipment into a registered
jack we received an alert and
are notified of its location.
Ports were bound to MAC addresses in order to prevent students or
passers-by from unplugging our computers and plugging in their own.
Not doing this would be a failure waiting to happen (see on MAC address
spoofing detection in \xs{sect:mac-spoofing-mon}).

A sample of a common switch port configuration in teaching and research labs
is in \xf{fig:typical-port-config}.

We used to \texttt{shutdown} ports automatically, requiring a manual
intervention to bring them back up by an analyst as a proactive way
to shut out
rogue clients, but it became evident
fast enough that this induced a significant manual intervention overhead
as most frequent violations were people moving their computers
from one desk to another,
or someone connecting their laptop from jack to jack in
a room until they realised it's not going to work.
The switch ports are configured to \api{restrict}.
The users do not get the network service until they talk to us or
to the service desk to notify of the fact that the equipment has moved
or new equipment was purchased and requires networking.
Switching ports to \api{protect} does silence the repeated
alerts without granting the network service until the computer
is moved back to its original jack or we are notified to authorise
the move.
It's worth noting \api{protect} has appeared with a particular
release of IOS, and prior to that \api{shutdown} was the only
option (needless to say the scale of manual intervention overhead required
to bring the ports back up). Additionally, while \api{restrict}
is a useful macro, it has a side effect of
complicating debugging when machines move by necessitating port clear-ups
from ``sticky'' MAC addresses. In the end, the user services\index{services} staff have been granted access to
that feature via a shell to clear up ports, etc. to address the management scalability\index{scalability}
problem.

\begin{figure}[htpb!]
\hrule
\begin{verbatim}

interface FastEthernet0/40
 description [Auto] machine.domain
 switchport access vlan XYZ
 switchport mode access
 switchport port-security
 switchport port-security violation restrict
 switchport port-security mac-address sticky
 switchport port-security mac-address sticky XXXX.XXXX.XXXX
 spanning-tree portfast
end
\end{verbatim}
\hrule
\normalsize
\caption{Typical port configuration for computers in labs}
\label{fig:typical-port-config}
\end{figure}

It is also worth mentioning the specific use of \api{spanning-tree portfast}
in light of the discussion in \xs{sect:spanning-tree}. This option
is used to speed up the (re)computation of the spanning tree helping
the overall network performance and nearly guaranteeing no loops \cite{cisco-kb-stp}.

\subsection{MAC spoofing monitoring}
\label{sect:mac-spoofing-mon}

MAC spoofing monitoring\index{monitoring!MAC spoofing} is a serious task to watch out for about
1000 Faculty-managed computers in teaching and general purpose labs
where physical presence even with a fleet of service desk personnel
is not always possible. Any student or visitor alike can unplug
a network cable from the wire and plug it into their laptop
having previously observed the MAC address of the analyst-managed
computer by logging to it first and then altering own's MAC address
of the laptop by a tool or a virtual environment such as VirtualBox, VMWare,
Parallels or the like. The desktops are expected to talk to us
in a certain way and the switch log has a sequence of link up/down
events when the jack is unplugged and plugged back in that
triggers the alert.
We had 3-5 incidents that we caught by the monitoring. There
are a few false positive cases usually caused by ghosting\index{ghosting} at
times or slow system startup. There is usually no swpvio, as
the MAC address is changed in advance on the offending laptop.

\subsection{Network bandwidth abuse}

We use Argus
\cite{argus,tao-net-sec-2005}
to monitor and audit aggregate
statistics about our hosts on the network on the routers. One of the scripts
from the scriptfarm\index{scriptfarm} (see \xs{sect:scripting}) reports regularly via a \tool{cron} job the top exporters and
importers of the data to and from the outside. Since we know who/what those
hosts are from our RP records in the database\index{database} (see \xs{sect:database}), we can follow up then with
the person via a ticketing system. In some cases exceptions are permitted
to known educational and patch sites alike
(see \xf{fig:typical-patch-sites})
or some research group's industrial partners. All other connections are ``choked''
to a limit to repay the bandwidth used (e.g. in \xf{fig:typical-choking-rules}).

\begin{figure}[htpb!]
\hrule
\scriptsize
\begin{verbatim}

# Note:  Ordered by netblock size for efficiency
17.0.0.0/8         Apple Computer
65.52.0.0/14       Microsoft
204.70.0.0/15      SAVVIS
131.107.0.0/16     Microsoft
207.46.0.0/16      Microsoft
64.4.0.0/18        Microsoft
207.68.128.0/18    Microsoft
80.231.19.64/27    Microsoft
198.6.32.0/19      Symantec
207.68.192.0/20    Microsoft
66.187.224.0/20    Red Hat
209.87.208.0/20    Zone Labs (zonealarm)
209.132.176.0/20   Red Hat
192.92.94.0/24     Symantec
216.10.192.0/24    Symantec
216.34.181.0/24    SourceForge
206.167.78.0/26    Akamai RISQ
213.86.172.128/27  Sophos
...
\end{verbatim}
\normalsize
\hrule
\caption{Typical patch sites}
\label{fig:typical-patch-sites}
\end{figure}

\begin{figure*}[htpb!]
\hrule
\scriptsize
\begin{verbatim}

Chain PREROUTING (policy ACCEPT)
target      prot opt source              destination
choke00044  all  --  anywhere            internal.host1.domain
choke00045  all  --  anywhere            internal.host2.domain
...
Chain POSTROUTING (policy ACCEPT)
target     prot opt source               destination
chokeOutsides  all  --  video.domain/19  anywhere
chokeOutsides  all  --  warez.domain/19  anywhere
...
Chain choke00044 (1 references)
target     prot opt source               destination
ACCEPT     all  --  anywhere             anywhere     limit: avg 20/sec burst 20
DROP       all  --  anywhere             anywhere

Chain choke00045 (1 references)
target     prot opt source               destination
ACCEPT     all  --  anywhere             anywhere     limit: avg 20/sec burst 20
DROP       all  --  anywhere             anywhere

Chain chokeOutsides (28 references)
target     prot opt source               destination
ACCEPT     all  --  anywhere             anywhere     limit: avg 100/sec burst 100
DROP       all  --  anywhere             anywhere
\end{verbatim}
\normalsize
\hrule
\caption{Typical choking rules}
\label{fig:typical-choking-rules}
\end{figure*}

\subsection{Nessus, Snort, and infection monitoring}
\index{monitoring!infections}
\index{monitoring!vulnerabilities}
\index{monitoring!tools}

Nessus\index{Tools!Nessus}, Snort\index{Tools!Snort} are typical tools
\cite{nessus,snort,linux-firewalls-ids-2007,tao-net-sec-2005}
we maintain at and run in our environment on all our clients. Critical problems are reported
to the RPs and any quarantine or blocking are instituted as necessary at the firewalls\index{firewall}.
We also monitor for infections, behaviour of port scanning, attempts to use outside
DNS servers other than ours, and the like, to which we respond and follow up by a variety of means.

\section{Management}
\label{sect:management}

Managing a network such as ours requires some knowledge
of the corresponding TCP/IP protocols \cite{tcp-ip-net-admin-93},
DNS \cite{dns-and-bind-98}, SNMP\index{SNMP}\index{Protocols!SNMP} \cite{stallings-snmp-99,essential-snmp-2001} and
the corresponding Linux networking \cite{linux-net-cookbook-2008}
to be able to write our tools that talk to switches, DNS,
ARP\index{ARP}\index{Protocols!ARP}, and others in accordance with the best practices
\cite{practice-sys-net-admin-99,net-management-2000}.

Scalability\index{scalability} is an issue.
During the course of the network's evolution, we went from
a few hundred computers and terminals running at 10~Mbps
to a few thousand at 100~Mbps, and a few dozen at 1~Gbps.
At present, downtime is approximately 30 seconds per month
(excepting individual switch failures) for monthly maintenance.
Delegating some network management responsibilities to the
service desk and desktop group alleviates the burden via custom
build scripts. We employed OSS tools that help us with that as
well and are in the process of adopting Netdisco\index{Tools!Netdisco}
\cite{netdisco}
with its web-based management console for the task to improve on
that end further.
Our overall topological perspective view of the switchfarm\index{switchfarm}
is illustrated in \xf{fig:netmap} from Netdisco. The labels are
intentionally obscured.
We use MRTG\index{Tools!MRTG}
\cite{mrtg,essential-snmp-2001}
as our graphing tool
to visualise the traffic flows on switch ports.

\begin{figure}[htbp!]
	\centering
	\includegraphics[width=.7\columnwidth]{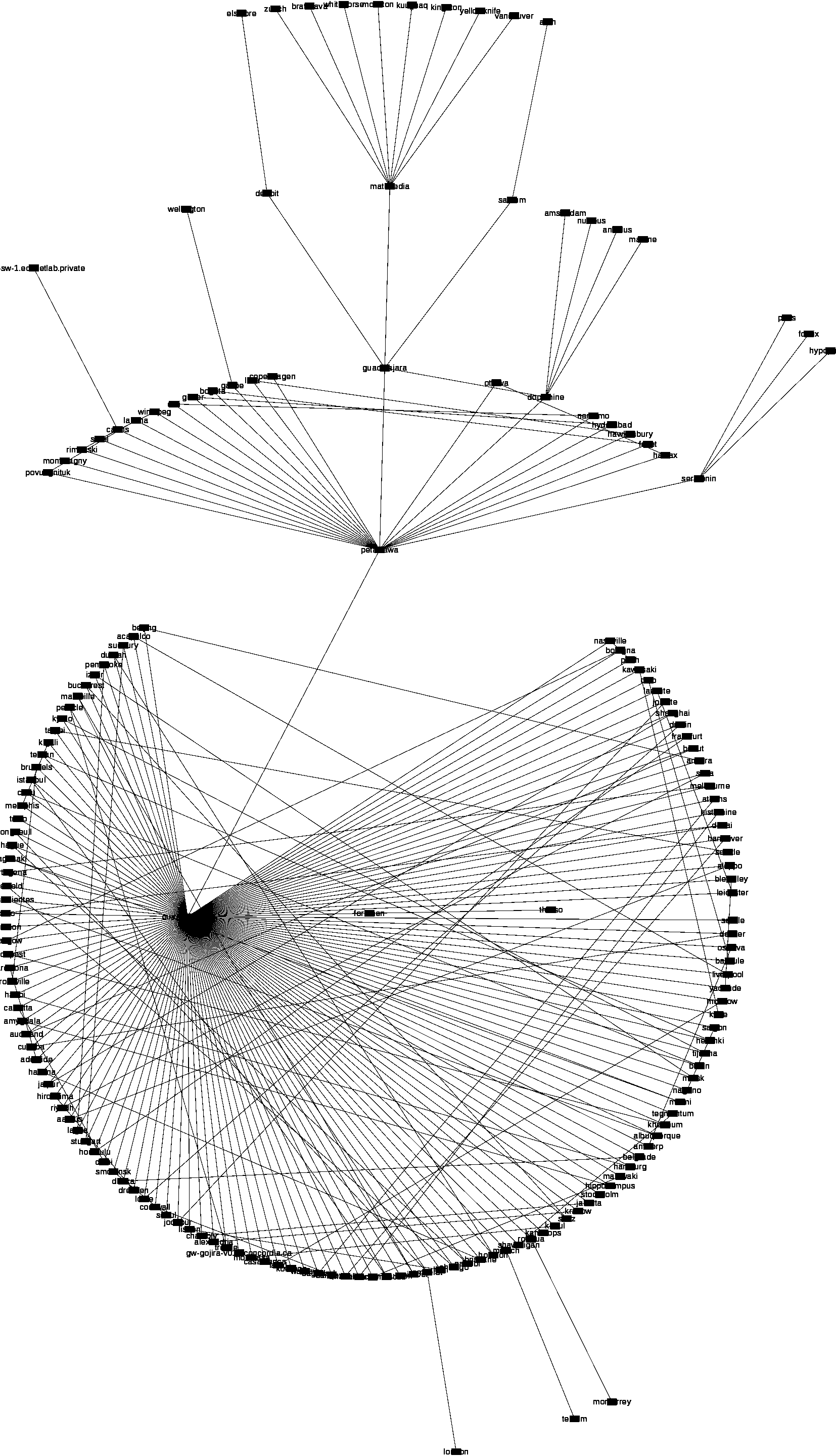}
	\caption{Netdisco's NetMap global switchfarm\index{switchfarm} overview}
	\label{fig:netmap}
\end{figure}

\subsection{Database}
\label{sect:database}
\index{database}

We designed our relatively large database\index{database} using a PostgreSQL 8.x\index{database!PostgreSQL}
\cite{postgres}
setup to record the information of
our switches, ports, patches, patch panels, jacks, rooms,
room occupants, RP records for hosts, user-managed and analyst-managed,
and the relationships between all those entities amounting at
present to about {\dbentities} relations (tables, views, triggers, constraints).
(See e.g. \xf{fig:dbdiag}).
Some of the information is maintained by scripts, such as what is connected where
and last seen on a switch or an ARP\index{ARP}\index{Protocols!ARP} table, list of VLANs from the VTP master,
and the like. The responsible person (RP) data are maintained manually as well as when new
switches are deployed or relocated and new jacks, patches, and ports are added.
A number of views
(the largely single rectangles on the image)
and interface scripts exists that allow
the networking group and sister groups to query some of the data to perform their job, specifically
by the service desk and the desktop groups for location awareness and default
services\index{services}, among others.

\begin{figure}[htbp!]
	\centering
	\includegraphics[width=\columnwidth]{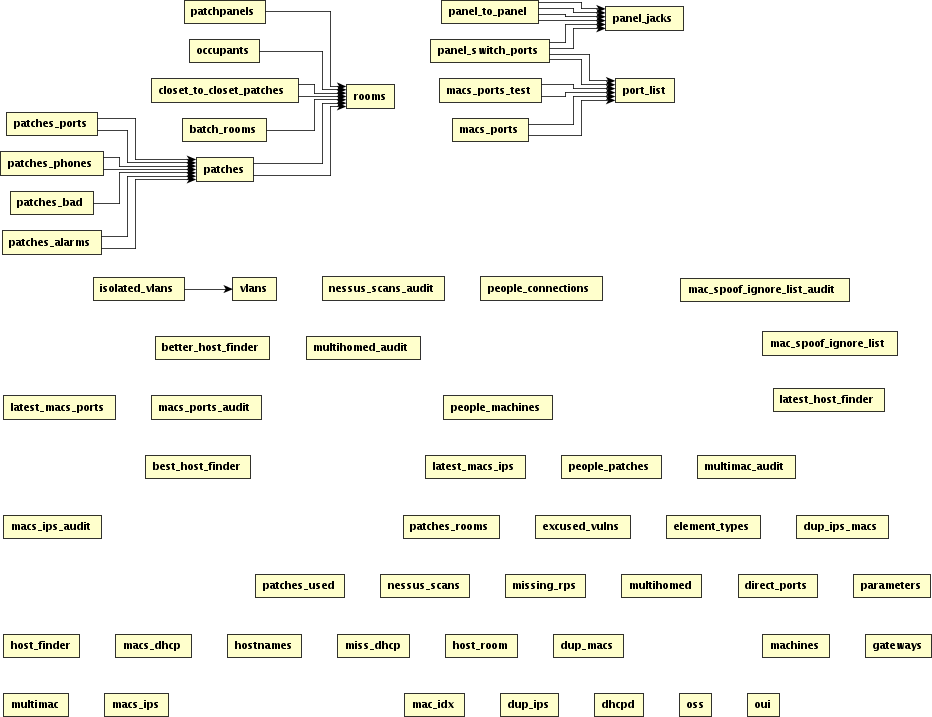}
	\caption{High-level database structure overview}
	\label{fig:dbdiag}
\end{figure}

\subsection{Scripting and automation}
\label{sect:scripting}

We have developed and continue developing, maintaining, refactoring what we
call our {\em scriptfarm}\index{scriptfarm} for the management of the switchfarm\index{switchfarm}, firewalls\index{firewall}, the database\index{database},
the switch logs, etc. The scripts also provide shells to other analysts
in sister groups to access the network data and in some cases change it
for us to scale. The majority of the scripts are written in Perl primarily
for its powerful regular expressions \cite{perl-regex-97},
set of libraries for network and database programming \cite{network-prog-with-perl-2001},
system administration \cite{perl-sysadmin-2000} and for
using Perl modules and objects \cite{perl-objects-2003}.
For some scripting we reply on Python \cite{core-python-prog-2007},
in particular for our Snort customizations, as well as shell scripts in \tool{tcsh} and \tool{bash}.

We maintain an SCM repository (CVS\index{Tools!CVS} \cite{cvs}), and central Makefile-based \cite{gmake}
build and deploy system, and the startup scripts we need to uniformly use on
our managed servers. Upon committing the revisions to the CVS the rest of the
members are notified. The updates are propagated through the CVS as well to the
server machines. The central consolidated repository for the all the numerous scripts in the scriptfarm\index{scriptfarm}
significantly improves maintenance and accounting of the scripts and code
auditing, and overall design and development process.

The automation aspect comes around maintaining the database about the current
state of the network as well as maintenance of the description fields on the
switch ports with a special \verb+[Auto]+ token set to make them more
usable and readable when viewing via MRTG\index{Tools!MRTG}
\cite{mrtg}
and other views and shells.
The automation of course covers security\index{security} alerts to either mailing lists, individuals,
or the RT ticketing system if something noteworthy happens for us to take action.
Switch configuration backup is also part of the scriptfarm\index{scriptfarm}.
We also build our startup init scripts, typically launched from
different run levels when the machine boots depending on which
cron job and other services\index{services} are assigned to a particular machine
at the boot time rather than to its back up peer.

\section{Conclusion}
\label{sect:conclusion}

We serve four different buildings with a network
segregated by VLANs. Our switching design uses
the spanning tree protocol\index{spanning tree}\index{Protocols!spanning tree} with redundancy.
Likewise we have redundant routing\index{routing} and firewalling\index{firewall} geographically spread.
We maintain a centralised DNS and DHCP.
We support ``multicast''\index{multicast} ghosting\index{ghosting} (not true multicasting just yet, but
rather broadcasting where the ghost data are transferred to every trunk).
We maintain in-house and employ external OSS management software and a database that controls network data
in a bidirectional manner.
Various security\index{security} scripts in the scriptfarm\index{scriptfarm} alert of problems.

Faced with IP address exhaustion we now have a motivational
drive to move to IPv6\index{IPv6}. Managing excessive aggregate bandwidth
is becoming a problem. Aging access layer and management software
is another issue we are facing. Some new challenges emerge
when building reorganizations take place. By the nature of our
work we will be collaborating with the other networking groups
beyond our faculty (e.g. access support for licensing, wireless,
and others), whose topology design is significantly different from ours
driving us to simplify operations. We are also dealing with
issues regarding increasing service requests, virtualisation support,
and maintaining data quality in our database.

Based on our experience, the broad guideline that we arrive at
is that it's useful to think about ways to let one's network evolve
over time rather than to redesign it from scratch whenever new
problems arise. Campus-wide VLANs\index{VLAN!Campus-wide} implementation is of particular
mention. We also proposed ways of using old ideas and old solutions
to existing and new problems.
The rest can be derived from our report and adapted
accordingly if needed for various aspects presented.

\subsection{Statistics}

\begin{itemize}
	\item 4 buildings served
 	\item about {\swTNFZ}, {\swTNXZ}, {\swTNSZ}, {\swTNTF}, {\swTFFZ}, {\swTFFE} switches
	\item about {\wclosets} wiring closets
	\item about {\netclients} clients
	\item about {\jackpatches} patched jacks
	\item about {\vlans} VLANs
	\item about {\scripts} scripts in the scriptfarm\index{scriptfarm}
	\item about {\dbentities} database relations (tables, views, triggers, etc.)
	\item
		about {\fwrulesIN} firewall\index{firewall} rules on the internal edge;
		{\fwrulesEX} firewall rules on the external edge
	\item network group members ranged between 2 and 4 (presently) in charge of
		the network in the Faculty.
\end{itemize}

\subsection{Future work}

As we are moving towards newer hardware and tools
and overall design, we are documenting the change
and plan on reporting an updated experience report
as a result. In summary, the future work in various
terms and durations will focus on:

\begin{enumerate}
	\item Gradual switchover from old to new core switches
	(see \xf{fig:during})
	and routers and a 10G backbone 
	eventually arriving at a topology shown in \xf{fig:after}%
.
	\item IPv6\index{IPv6} pilot project
\cite{migrating-to-ipv6-2006,theipv6experts,nist-ipv6,implementing-ipv6-cuv-2007}
	the first in the University.
	\item Layers 2\index{switchfarm!Layer 2} and 3\index{switchfarm!Layer 3} switching / routing\index{routing} with eventual
	switchover of the internal router/firewall\index{firewall} to the {\cisco} \catalyst{4510} and \catalyst{3750X} switches.
	\item
		{\cisco} ACLs to reduce/eliminate Linux internal routers' \tool{iptables} load.
		To achieve that our ongoing preliminary plan is to simplify, consolidate, and eliminate
		some of the complexity of the current ruleset before actually moving onto the ACLs.
	\item Consolidate scripting with standardization and refactoring.
	\item Cloud/mobility/ubiquity support of various networking equipment.
	\item Network preparation for Active Directory (AD) service.
	\item Explore the OpenNMS\index{Tools!OpenNMS} \cite{opennms} for a variety of management tasks in
	addition to our existing toolset.
\end{enumerate}

\section*{Acknowledgment}

This work is supported by the Faculty of Engineering and Computer Science,
Concordia University, Montreal, QC, Canada.
We acknowledge the feedback, bug reports,
and suggestions from our sister IT groups
of System Administration Group, Desktop
Operations Group, User Services Group, Faculty Information Systems group,
techies,
and other
colleagues.

\bibliographystyle{plain}
\bibliography{encs-networking-arxiv}

\printindex

\end{document}